\documentclass[preprint,5p,times,twocolumn]{elsarticle}
\usepackage{lineno,hyperref}
\usepackage{hyperref}
\usepackage{subcaption}
\usepackage{xcolor}
\usepackage{soul}

\usepackage{graphicx,color,amsmath,amssymb,chemformula}

\begin{document}
\begin{frontmatter}
\title{Geometry-controlled
magnon-polaritons of double magnetic films in planar cavities}

\author{S Solihin$^{\text{a,b,}*}$}
\ead{solihin@ui.ac.id}

\author{Ahmad R. T. Nugraha$^{\text{b,c}}$}
\ead{ahma080@brin.go.id}

\author{Muhammad Aziz Majidi$^{\text{a}}$}
\ead{aziz.majidi@sci.ui.ac.id}

\address{
\vspace{1mm}
$^{a}$Department of Physics, Faculty of Mathematics and Natural Sciences, Universitas Indonesia, Depok 16424, Indonesia\\
$^{b}$Research Center for Quantum Physics, National Research and Innovation Agency (BRIN), South Tangerang 15314, Indonesia\\
$^{c}$Engineering Physics Study Program, School of Electrical Engineering, Telkom University, Bandung 40257, Indonesia
}
\begin{abstract}
Planar cavity magnonics has been developed mainly for a single magnetic film, leaving multilayer behavior in spatially resolved cavity scattering largely unexplored. Here, we introduce a double layer planar cavity with two magnetic films embedded in the same microwave cavity to derive a full two-film scattering theory in the macrospin ($J = 0$) limit and recover the exact zero-gap half-thickness limit, thereby benchmarking the model against the known one-film result. We find that the double layer model actively enables geometry-controlled bright-channel enhancement, demonstrating that the magnon-photon coupling depends on spatial placement rather than just total magnetic volume. Antinode-compatible placements increase the coupling, while node-compatible placements suppress it. Weak symmetry breaking also transfers finite cavity weight to a mode dark in the symmetric limit, producing an additional branch without destroying the main avoided crossing. Finally, a reduced multimode theory for $J\neq 0$ predicts family-resolved bright and dark channels for odd standing-spin-wave modes.
\end{abstract}

\end{frontmatter}

\section{Introduction}
\label{sec:introduction}

Magnonics is the study of the generation, propagation, control, and detection of spin waves and their quanta in magnetically ordered media \cite{rezende2020fundamentals,Kruglyak2010Magnonics, Lenk2011BuildingBlocks, ZareRameshti2022CavityMagnonics}. This field has emerged as a promising wave-based information platform because magnons can encode both phase and amplitude \cite{Schneider2008SpinWaveLogic, Chumak2014MagnonTransistor, Barman2021Roadmap}. Furthermore, magnons can be guided and interfered with on submicron scales \cite{Chumak2017MagnonicCrystals, Flebus2024Roadmap}, and can be interfaced naturally with microwave \cite{Demokritov2006BECMagnons, bittencourt2025magnon}, spintronic \cite{Chumak2015MagnonSpintronics,Cornelissen2015LongDistance}, and hybrid quantum architectures \cite{rusconi2019hybrid, muneeb2026magnon}. Over the past two decades, this perspective has evolved from demonstrations of wave-based logic and transport to a broader program in which magnonic functionality is engineered through geometry \cite{devapriya2026lattice, myhre2026thickness, Krawczyk2014MagnonicCrystals}, confinement \cite{chen2024magnon, chen2026magnon, husain2026magnon}, mode hybridization \cite{mamica2026impact,li2024magnon}, and materials design \cite{zhang2023review, han2024magnonics}. In this context, ferrimagnetic insulators such as yttrium iron garnet (YIG) have played a central role due to their exceptionally low damping and ability to support both nearly uniform and strongly nonuniform spin-wave modes \cite{Kajiwara2010Transmission, Serga2010YIGMagnonics}.

A major branch of this broader effort is cavity magnonics, in which magnons coherently hybridize with confined electromagnetic modes \cite{harder2018cavity,ZareRameshti2022CavityMagnonics}. Early theory established that magnetic collective excitations can couple strongly to cavity photons \cite{Soykal2010NanomagnetCavity}, and some follow-up experiments immediately demonstrated strong and ultrastrong magnon-photon coupling in superconducting resonators \cite{Huebl2013HighCooperativity, Goryachev2014HighCooperativity, ghirri2023ultrastrong, ghirri2024interplay}, three-dimensional microwave cavities \cite{Tabuchi2014QuantumLimit, Zhang2014StronglyCoupled}, and layered systems \cite{golovchanskiy2021ultrastrong, hong2026strong}. These developments were later extended to spin pumping \cite{Bai2015SpinPumping, maier2016spin}, qubit-based hybrid architectures \cite{Tabuchi2015Qubit, Yuan2022QuantumMagnonics}, and other hybrid quantum platforms \cite{Awschalom2021HybridMagnonics, LachanceQuirion2019HybridQuantum}. More recently, cavity magnonics has expanded to include non-Hermitian effects \cite{Harder2018LevelAttraction, Wang2019Nonreciprocity}, interferometric control \cite{Rao2021PerfectAbsorption, zhao2021phase}, topology \cite{Lee2023Topological, hirosawa2022magnetoelectric}, and shape-engineered coupling profiles \cite{MartinezLosa2023ShapedFerromagnets, shuai2025precise}, demonstrating that the structure of magnon-photon hybridization can be engineered beyond the simplest avoided-crossing picture.

A particularly groundbreaking theoretical work in cavity magnonics is the planar scattering theory developed by Cao~\textit{et al.} \cite{Cao2015ExchangeMagnonPolaritons}. They formulated magnon-photon coupling directly at the level of coupled Maxwell and Landau-Lifshitz-Gilbert (LLG) equations, rather than starting from an independent-spin or rotating-wave approximation. The theory naturally captures cavity loading, thickness dependence, and the crossover from a macrospin-like ferromagnetic resonance to exchange-driven standing-spin-wave resonances within a unified planar-cavity framework. This point is especially important in planar geometries, where the cavity mode profile, dielectric loading, and sample placement can all substantially modify the transmission spectrum. In this sense, their theory provides a strong starting point for extending planar cavity magnonics beyond the single-film geometry~\cite{ZareRameshti2022CavityMagnonics}.

Ingredients closely related to this problem have already been reported in the literature. It is widely recognized that the coupling of multiple macroscopic magnons to the same cavity mode has been extensively studied, particularly for applications involving entanglement generation \cite{baghshahi2025generation,li2019entangling} and squeezing in quantum magnonics \cite{qi2023magnon, weng2026magnon}. In parallel, bright and dark collective magnon modes, together with the characteristic collective enhancement of the bright channel, were demonstrated for two YIG spheres coupled to a microwave cavity by Zhang \textit{et al.} \cite{Zhang2015DarkModes}. Multiple magnetic objects coupled through a common cavity have also been studied in other forms, including cavity-mediated dissipative spin-spin coupling for two YIG films \cite{Grigoryan2019DissipativeSpinSpin}, planar-cavity dark-mode spin pumping in a Pt/YIG-based metamaterial structure \cite{Pan2023SpinPumpingDarkMode}, and bright and dark magnon modes in an exchange-coupled ferromagnetic double layer in a microwave cavity \cite{Zhan2021BrightDarkbilayers}. However, relying on generic coupled-oscillator models for two magnons is fundamentally insufficient for spatially extended planar systems. Therefore, such developments do not remove the need for a double layer planar full-scattering theory because the combination of planar cavity walls, finite spacer regions, standing-wave placement, zero-gap consistency, and exchange-sector multimode structure remains only partially explored within a single geometry-resolved framework \cite{Cao2015ExchangeMagnonPolaritons,ZareRameshti2022CavityMagnonics,Yuan2022QuantumMagnonics}. Such a gap of exploration is timely because current magnonics roadmaps emphasize reconfigurable wave control, collective-mode engineering, and hybrid functionality across classical and quantum regimes \cite{Barman2021Roadmap,Flebus2024Roadmap}. In parallel, recent cavity-magnonics studies have highlighted the importance of mode overlap, cavity interference, ferromagnet shape, and higher-order magnetic resonances for controlling hybrid spectra \cite{Rao2021PerfectAbsorption,MartinezLosa2023ShapedFerromagnets,Smith2024ExchangeCMP,Hoshi2024StandingSpinWaves,Zhuang2024HybridMagnonPhonon}. These observations establish the double layer planar cavity as a fundamentally distinct system beyond the single-film problem. Within a spatially explicit cavity-scattering theory, several questions naturally arise concerning how the cavity standing-wave pattern controls collective bright enhancement, how controlled asymmetry activates weak dark channels, and how exchange-driven standing-spin-wave families reorganize in a double layer environment \cite{ZareRameshti2022CavityMagnonics,Smith2024ExchangeCMP,Hoshi2024StandingSpinWaves,Zhuang2024HybridMagnonPhonon}.

In this work, we comprehensively develop the double layer planar cavity theory in two stages. First, a double layer extension of the planar single-film scattering theory in the macrospin limit is constructed and validated. The resulting framework reproduces the single-film cavity transmission exactly in the zero-gap half-thickness limit and reveals a geometry-controlled bright-channel enhancement, including a $\sqrt{2}$-type increase of the effective coupling at antinode-compatible placements and a coupling suppression at node-compatible placements. Second, we break the double layer symmetry and show that an additional dark branch in the ideal symmetric limit becomes weakly visible in transmission while the main bright splitting remains substantial. Motivated by the exchange-sector results for the single-film planar cavity~\cite{Cao2015ExchangeMagnonPolaritons}, we then formulate a reduced multimode double layer theory for $J\neq 0$, in which each odd standing-spin-wave family develops its own bright and dark double layer channels. Taken together, these results indicate that double layer planar cavity magnonics is a useful route toward geometry-controlled collective-mode engineering beyond the single-film planar cavity \cite{Cao2015ExchangeMagnonPolaritons,Zhang2015DarkModes,ZareRameshti2022CavityMagnonics,Yuan2022QuantumMagnonics,Zhan2021BrightDarkbilayers}.

\section{Model and theoretical framework}
\label{sec:model}

In this section, we introduce the theoretical framework used throughout the paper, which extends the single-film planar cavity scattering theory developed by Cao et al. \cite{Cao2015ExchangeMagnonPolaritons} to a double layer geometry. We first formulate the full double layer scattering theory in the macrospin limit ($J=0$) and establish its validity by recovering the known single-film limit. After that, we introduce the reduced exchange-double layer multimode theory for $J\neq 0$, which extends the same physical picture to standing-spin-wave resonances and family-resolved bright and dark double layer channels.

\begin{figure}[t]
    \centering
    \includegraphics[width=1\linewidth]{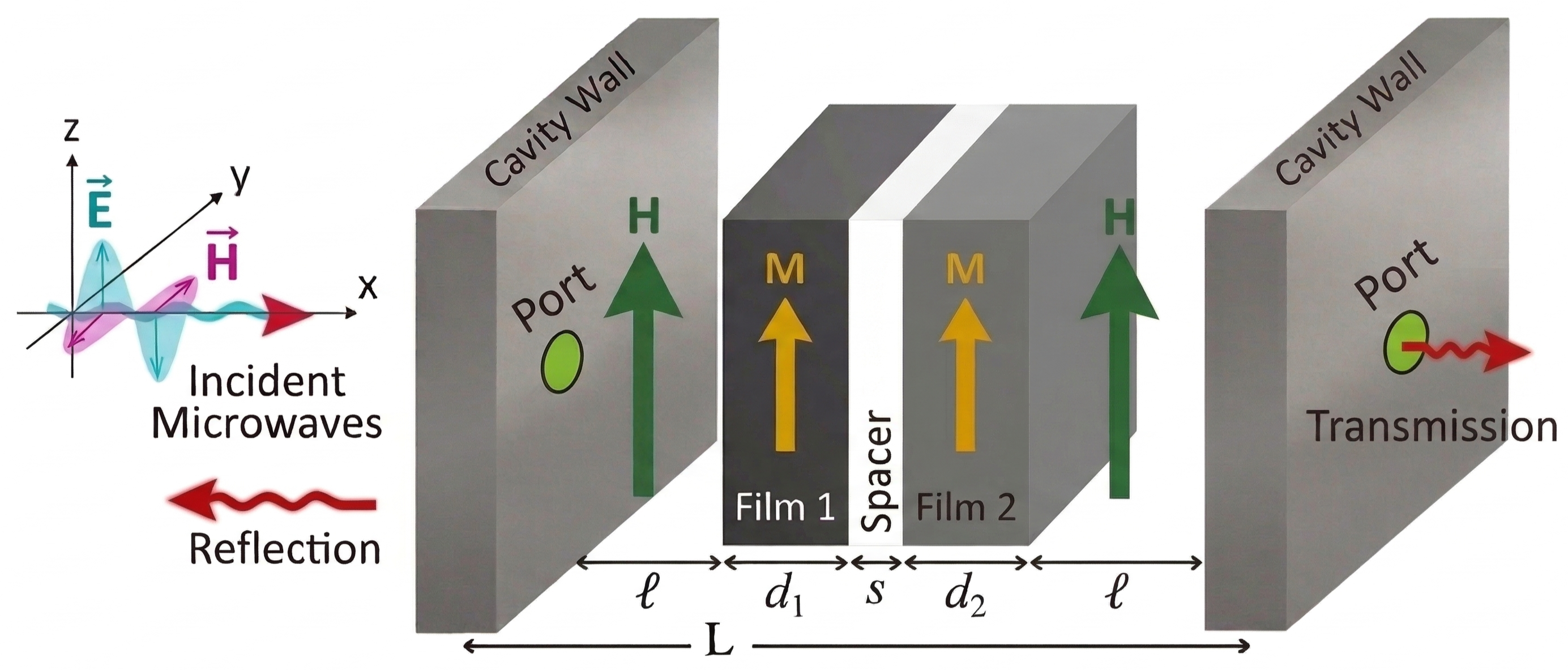}
    \caption{Double magnetic film in a planar electromagnetic cavity.}
    \label{fig:ilustration}
\end{figure}
\subsection{Single-film benchmark and double layer planar cavity geometry}
\label{subsec:geometry}

Let us first consider a one-dimensional planar microwave cavity of total length $L$, bounded by two partially transmitting cavity walls located at $x=0$ and $x=L$. This geometry follows the planar cavity setting introduced in the single-film scattering theory of Ref.~\cite{Cao2015ExchangeMagnonPolaritons}, which we will refer throughout this work as the fundamental benchmark. The cavity axis is taken along the $x$ direction, while the static bias field is applied along a fixed transverse direction such that the microwave field couples to the magnetic dynamics of the film. As in Ref.~\cite{Cao2015ExchangeMagnonPolaritons}, we assume harmonic time dependence of the form $e^{-i\omega t}$ so that the problem reduces to a frequency-domain scattering calculation at angular frequency $\omega$.

The single-film benchmark corresponds to a magnetic film of thickness $d$ placed inside the cavity and separated from the two cavity walls by nonmagnetic spacer regions. In the symmetric reference configuration, the magnetic slab is centered inside the cavity, so that the left and right spacer lengths are equal:
\begin{equation}
\ell_{\mathrm{sf}}=\frac{L-d}{2}.
\label{eq:singlefilm_spacer}
\end{equation}

The double layer geometry considered here is the minimal extension of that single-film planar cavity. As shown in Figure~\ref{fig:ilustration}, two magnetic films of thicknesses $d_1$ and $d_2$ are inserted into the same cavity and separated by a nonmagnetic internal spacer of length $s$. The full scattering problem thus consists of seven spatial regions comprising the two exterior regions, the two cavity spacers, the internal nonmagnetic spacer, and the two magnetic films. We focus first on the mirror-symmetric placement of the two films inside the cavity, for which the left and right outer spacer lengths are equal and denoted by $\ell$. Simple geometric consistency then gives
\begin{equation}
\ell=\frac{L-d_1-d_2-s}{2}.
\label{eq:double layer_outer_spacer}
\end{equation}
Equation~\eqref{eq:double layer_outer_spacer} makes clear that the double layer geometry is controlled by three independent structural parameters, namely the two film thicknesses $d_1$ and $d_2$, and the inter-film separation $s$.

When the two films are identical ($d_1=d_2=d$), the film centers are located at
\begin{align}
x_1&=\ell+\frac{d}{2}
=\frac{L-d-s}{2},\\
x_2&=L-\ell-\frac{d}{2}
=\frac{L+d+s}{2}.
\label{eq:film_centers_equal_thickness}
\end{align}
These positions play an important physical role because they determine whether the two films sit near cavity antinodes or cavity nodes of a given standing-wave mode. As a result, the same total magnetic material can produce very different effective couplings depending on the double layer placement within the cavity.

The cavity walls are modeled, as in the benchmark planar theory, as partially transmitting interfaces characterized by a single opacity parameter $\Delta$~\cite{Cao2015ExchangeMagnonPolaritons}. A crucial consistency requirement follows immediately from the relation between the one-film and two-film geometries. If a single benchmark film of thickness $d$ is divided into two identical half-thickness films,
\begin{equation}
d_1=d_2=\frac{d}{2},
\qquad
s=0,
\label{eq:half_half_geometry_condition}
\end{equation}
then the double layer should collapse to the original one-film configuration. At the geometric level, Eq.~\eqref{eq:half_half_geometry_condition} gives
\begin{equation}
\ell=\frac{L-d}{2},
\label{eq:collapsed_outer_spacer}
\end{equation}
which naturally recovers the single-film spacer length in Eq.~\eqref{eq:singlefilm_spacer}.

\subsection{Full double layer scattering theory in the macrospin limit \texorpdfstring{$J=0$}{J=0}}
\label{subsec:j0_scattering}

We now construct the full double layer cavity-scattering theory in the macrospin limit. Generalizing the single-film framework \cite{Cao2015ExchangeMagnonPolaritons}, we consider two magnetic slabs separated by a finite nonmagnetic spacer and embedded inside a cavity bounded by partially transmitting walls with opacity $\Delta$.

Assuming a harmonic time dependence $e^{-i\omega t}$, the stationary scattering problem at angular frequency $\omega$ involves the vacuum microwave wave number $q=\omega/c$. Inside a magnetic film with relative permittivity $\eta$, the dielectric wave number is $k_{\varepsilon}=\sqrt{\eta}\,q$. In the macrospin limit ($J=0$), the magnetic response modifies the internal propagation through the effective film wave number
\begin{equation}
k(\omega,H)
=
k_{\varepsilon}
\sqrt{
1+u-\frac{v^2}{1+u}
},
\label{eq:k_macrospin_main}
\end{equation}
with
\begin{equation}
u=\frac{\omega_k\omega_M}{\omega_k^2-\omega^2},
\qquad
v=\frac{\omega\omega_M}{\omega_k^2-\omega^2},
\qquad
\omega_k=\omega_H-i\alpha\omega.
\label{eq:u_v_main}
\end{equation}
Here, $\omega_H=\gamma\mu_0 H$ and $\omega_M=\gamma\mu_0 M_s$, where $\gamma$ is the gyromagnetic ratio, $H$ is the external bias field, $M_s$ is the saturation magnetization, and $\alpha$ is the Gilbert damping constant \cite{Cao2015ExchangeMagnonPolaritons}. Equation~\eqref{eq:k_macrospin_main} establishes the fundamental cavity-magnon anticrossings at ferromagnetic resonance (FMR) mode $\omega_{c,n} = \omega_{\text{FMR}}$, yielding the resonance magnetic field \cite{Cao2015ExchangeMagnonPolaritons}:
\begin{equation}
\mu_0 H_{\text{res},n} = \frac{-\omega_M +\sqrt{\omega_M^2+4\omega_{c,n}^2}}{2\gamma}.
\label{eq:H_res_main}
\end{equation}

The cavity walls are modeled by an opacity parameter $\Delta$. Their transmission and reflection amplitudes are
\begin{equation}
t_c=\frac{i}{i+q\Delta},
\qquad
r_c=-\frac{q\Delta}{i+q\Delta}.
\label{eq:tc_rc_main}
\end{equation}
At a cavity-film interface, the effective impedance mismatch is governed by
\begin{equation}
\beta=\frac{\eta q-k}{\eta q+k}.
\label{eq:beta_main}
\end{equation}

For later comparison, the benchmark single-film cavity transmission amplitude of thickness $d$ is \cite{Cao2015ExchangeMagnonPolaritons}:
\begin{equation}
S_{\text{single}}(\omega,H)
=
\frac{(1-\beta^2)t_c^2 e^{i(k-q)d}}
{\left(1-\beta r_c e^{i\phi}\right)^2
-
e^{2ikd}\left(\beta-r_c e^{i\phi}\right)^2},
\label{eq:Ssingle_eq9_main}
\end{equation}
where $\phi=q(L-d)$. Isolating the magnetic slab as a single scatterer embedded in the cavity medium, its exact reflection and transmission amplitudes are
\begin{equation}
r_{\mathrm{slab}}
=
\frac{\beta\left(1-e^{2ikd}\right)}
{1-\beta^2 e^{2ikd}},
\qquad
t_{\mathrm{slab}}
=
\frac{(1-\beta^2)e^{ikd}}
{1-\beta^2 e^{2ikd}}.
\label{eq:rt_slab_main}
\end{equation}
It is important to clarify the phase convention used in Eq. \ref{eq:rt_slab_main}. At the isolated-slab level, the transmission phase is $e^{ikd}$, representing the phase accumulation over the thickness $d$ for a wave propagating along the positive $x$-axis. Similarly, the term $e^{2ikd}$ in the reflection coefficient accounts for the coherent round-trip phase delay acquired inside the magnetic medium. By contrast, the factor $e^{i(k-q)d}$ in Eq. \ref{eq:Ssingle_eq9_main} belongs to the already reorganized one-film cavity expression. This distinction is essential for obtaining the correct double layer theory and for recovering the single-film result in the zero-spacing limit.

For the double layer configuration, let film $\nu$ ($\nu=1,2$) have thickness $d_\nu$, wave number $k_\nu$, mismatch parameter $\beta_\nu$, and isolated-slab amplitudes
\begin{equation}
r_{\nu}
=
\frac{\beta_{\nu}\left(1-e^{2ik_{\nu}d_{\nu}}\right)}
{1-\beta_{\nu}^2 e^{2ik_{\nu}d_{\nu}}},
\qquad
t_{\nu}
=
\frac{(1-\beta_{\nu}^2)e^{ik_{\nu}d_{\nu}}}
{1-\beta_{\nu}^2 e^{2ik_{\nu}d_{\nu}}}.
\label{eq:r_t_eachfilm_main}
\end{equation}
Summing the multiple reflections across the spacer $s$, the effective double layer transmission amplitude is
\begin{equation}
t_{\mathrm{eff}}
=
\frac{t_1 t_2 e^{iqs}}
{1-r_1 r_2 e^{2iqs}},
\label{eq:t_eff_main}
\end{equation}
and the effective reflection amplitudes for waves incident from the left and right are
\begin{equation}
r_{\mathrm{eff}}^{L}
=
r_1+\frac{t_1^{\,2} r_2 e^{2iqs}}
{1-r_1 r_2 e^{2iqs}},
\qquad
r_{\mathrm{eff}}^{R}
=
r_2+\frac{t_2^{\,2} r_1 e^{2iqs}}
{1-r_1 r_2 e^{2iqs}}.
\label{eq:r_eff_left_right_main}
\end{equation}
For identical films ($d_1=d_2=d, k_1=k_2=k, \beta_1=\beta_2=\beta, r_1=r_2=r, t_1=t_2=t$), these reduce to $r_{\mathrm{eff}}^{L}=r_{\mathrm{eff}}^{R} \equiv r_{\mathrm{eff}}$, restoring left-right symmetry.

To insert the effective double layer back into the cavity, we utilize a transfer-matrix representation \cite{nulli2018significant, kengo2024tunable}. For a reciprocal but generally asymmetric scatterer with left reflection amplitude $r^L$, right reflection amplitude $r^R$, and transmission amplitude $t$, the transfer matrix is written as
\begin{equation}
M_{\mathrm{asym}}(r^L,r^R,t)
=
\frac{1}{t}
\begin{pmatrix}
t^2-r^L r^R & r^R \\
-r^L & 1
\end{pmatrix}.
\label{eq:Masym_main}
\end{equation}
When the scatterer is left-right symmetric, this form reduces to
\begin{equation}
M_{\mathrm{sym}}(r,t) = \frac{1}{t}
\begin{pmatrix}
t^2-r^2 & r \\
-r & 1
\end{pmatrix}.
\label{eq:Msym_main}
\end{equation}
The free propagation matrix through a cavity spacer of length $\ell$ is defined as
\begin{equation}
P(q,\ell)
=
\begin{pmatrix}
e^{iq\ell} & 0 \\
0 & e^{-iq\ell}
\end{pmatrix}.
\label{eq:P_main}
\end{equation}

Because the cavity walls are symmetric scatterers, each wall is explicitly represented by
\begin{equation}
M_{\mathrm{wall}}
=
M_{\mathrm{sym}}(r_c,t_c)
=
\frac{1}{t_c}
\begin{pmatrix}
t_c^2-r_c^2 & r_c \\
-r_c & 1
\end{pmatrix}.
\label{eq:Mwall_main}
\end{equation}
The double layer itself is represented by an effective transfer matrix $M_{\mathrm{eff}}$. In the general asymmetric case, this matrix is constructed using the asymmetric form
\begin{equation}
M_{\mathrm{eff}}
=
M_{\mathrm{asym}}\!\left(r_{\mathrm{eff}}^{L},r_{\mathrm{eff}}^{R},t_{\mathrm{eff}}\right).
\label{eq:Meff_main}
\end{equation}
However, if the two magnetic films are perfectly identical and symmetric, the effective amplitudes satisfy $r_{\mathrm{eff}}^{L}=r_{\mathrm{eff}}^{R}=r_{\mathrm{eff}}$, allowing us to instead use the symmetric form $M_{\mathrm{eff}} = M_{\mathrm{sym}}(r_{\mathrm{eff}},t_{\mathrm{eff}})$.

Using these components, the total transfer matrix of the full double layer cavity becomes
\begin{equation}
M_{\mathrm{tot}}
=
M_{\mathrm{wall}}
\,P(q,\ell)\,
M_{\mathrm{eff}}
\,P(q,\ell)\,
M_{\mathrm{wall}},
\label{eq:Mtot_main}
\end{equation}
where $\ell=(L-d_1-d_2-s)/2$ represents the length of each exterior cavity spacer. The cavity transmission amplitude is then extracted from the total transfer matrix as
\begin{equation}
S_{\mathrm{double}}(\omega,H)
=
\frac{\det M_{\mathrm{tot}}}{\left(M_{\mathrm{tot}}\right)_{22}}.
\label{eq:Sbi_main}
\end{equation}

This formulation immediately recovers an important internal consistency requirement. If the original single film of thickness $d$ is divided into two identical half-thickness layers with zero gap ($d_1=d_2=d/2$ and $s=0$), the effective double layer amplitudes perfectly reduce to the single-slab values $r_{\mathrm{eff}}^{L}=r_{\mathrm{eff}}^{R}=r_{\mathrm{slab}}(d)$ and $t_{\mathrm{eff}}=t_{\mathrm{slab}}(d)$. Consequently, the double layer transmission exactly reproduces the single-film result, yielding
\begin{equation}
S_{\mathrm{double}}(\omega,H;d/2,d/2,s=0)
=
S_{\mathrm{single}}(\omega,H;d).
\label{eq:zero_gap_identity_main}
\end{equation}

\subsection{Reduced double layer multimode theory for \texorpdfstring{$J \neq 0$}{J ≠ 0}}
\label{subsec:exchange_reduced}

Having established the exact $J=0$ theory, we now formulate a reduced exchange-sector extension to analyze standing-spin-wave magnon-polariton spectra. Rather than constructing the full exact seven-region exchange solver, we derive a controlled reduced multimode theory that generalizes to a two-film geometry while retaining the essential exchange-induced physics. These include multiple internal magnetic branches, pinned-boundary standing-spin-wave resonances, and mode-dependent cavity couplings \cite{Cao2015ExchangeMagnonPolaritons}.

The key distinction is that the effective magnetic field now contains the exchange contribution $\mathbf{H}_{\mathrm{ex}} = J \nabla^2 \mathbf{m}$ \cite{Cao2015ExchangeMagnonPolaritons}. This wave-vector dependence produces a ladder of standing-spin-wave resonances. For a single pinned-boundary film of thickness $d$, the standing-spin-wave resonances (SWR) occur at \cite{Cao2015ExchangeMagnonPolaritons}:
\begin{equation}
\omega_{\mathrm{SWR}}^{(p)}
=
\sqrt{
\left[
\omega_H + 2J\omega_M\left(\frac{p\pi}{d}\right)^2
\right]
\left[
\omega_M + \omega_H + 2J\omega_M\left(\frac{p\pi}{d}\right)^2
\right]
},
\label{eq:omega_swr_single}
\end{equation}
for $p=1,2,3,\dots$ with $\omega_H=\gamma\mu_0 H$ and $\omega_M=\gamma\mu_0 M_s$. The cavity-magnon anticrossings condition, $\omega_{c,n} = \omega_{\text{SWR}}^{(p)}$, yields the resonance magnetic field \cite{Cao2015ExchangeMagnonPolaritons}:
\begin{equation}
\mu_0 H_{\text{res},n}^{(p)} \simeq \frac{-\omega_M - 2J\omega_M \left(\frac{p\pi}{d}\right)^2 +\sqrt{\omega_M^2+4\omega_{c,n}^2}}{2\gamma}.
\label{eq:H_res_swr}
\end{equation}

Within the standard input-output formalism \cite{tavis1968exact,walls2008quantum}, the transmission amplitude for a given odd cavity mode $n$ is:
\begin{equation}
S_{n,\text{single}}(\omega)
=
\frac{\kappa_{c,n}}
{i(\omega-\omega_{c,n})-\kappa_{c,n}
-
i\displaystyle\sum_{p\in\mathcal O}
\frac{\bigl(g_{n}^{(p)}\bigr)^2}
{\omega-\omega_{\mathrm{SWR}}^{(p)}+i\kappa_{n}^{(p)}}},
\label{eq:singlefilm_exchange_reduced_main}
\end{equation}
where $\omega_{c,n}$ and $\kappa_{c,n}$ are the cavity frequency and linewidth, $\omega_{\mathrm{SWR}}^{(p)}$ is the $p$-th standing-spin-wave resonance, $\kappa_n^{(p)}$ is the damping rate, and $g_n^{(p)}$ is the coupling strength \cite{harder2018cavity}. Regarding the symmetry, in both the centered single layer and the symmetrically placed double-film configurations, only odd standing-wave indices are expected to be appreciably visible \cite{Cao2015ExchangeMagnonPolaritons}, hence $p \in \mathcal{O} = \{1, 3, 5, \dots\}$.

Generalizing Eq.~\eqref{eq:singlefilm_exchange_reduced_main} to a double layer, let film $\nu$ ($\nu=1,2$) have parameters $d_\nu$, $J_\nu$, $M_{s,\nu}$, and $H_\nu$. The standing-spin-wave ladder is:
\begin{align}
\omega_{\mathrm{SWR},\nu}^{(p)}
=&
\left[
\omega_{H,\nu} + 2J_\nu\omega_{M,\nu}\left(\frac{p\pi}{d_\nu}\right)^2
\right]^{1/2}\nonumber\\
&\times\left[
\omega_{M,\nu} + \omega_{H,\nu} + 2J_\nu\omega_{M,\nu}\left(\frac{p\pi}{d_\nu}\right)^2
\right]^{1/2}
,
\label{eq:omega_swr_double layer_eachfilm}
\end{align}
for $p\in\mathcal O$ with $\omega_{H,\nu}=\gamma_\nu \mu_0 H_\nu$ and $\omega_{M,\nu}=\gamma_\nu \mu_0 M_{s,\nu}$.

If the films interact only via the cavity, the self-energy is additive:
\begin{equation}
\Sigma_{n,\mathrm{double}}^{0}(\omega)
=
\sum_{p\in\mathcal O}
\left[
\frac{\bigl(g_{n,1}^{(p)}\bigr)^2}
{\omega-\omega_{\mathrm{SWR},1}^{(p)}+i\kappa_{n,1}^{(p)}}
+
\frac{\bigl(g_{n,2}^{(p)}\bigr)^2}
{\omega-\omega_{\mathrm{SWR},2}^{(p)}+i\kappa_{n,2}^{(p)}}
\right].
\label{eq:sigma_double layer_additive}
\end{equation}
Although Eq.~\eqref{eq:sigma_double layer_additive} provides a useful first approximation, it cannot describe the collective bright and dark reorganization of each family, as direct inter-film interaction is omitted. To achieve this, the two branches of a given family $p$ must be allowed to hybridize.

We introduce an effective $2\times 2$ magnetic mode matrix $\Omega_p$, damping matrix $K_p$, and coupling vector $\mathbf{g}_p$:
\begin{equation}
\Omega_p = \begin{pmatrix}
\omega_{\mathrm{SWR},1}^{(p)} & J^{(p)}_{\mathrm{int}} \\
J^{(p)}_{\mathrm{int}} & \omega_{\mathrm{SWR},2}^{(p)}
\end{pmatrix},\quad
K_p = \begin{pmatrix}
\kappa_{n,1}^{(p)} & 0 \\
0 & \kappa_{n,2}^{(p)}
\end{pmatrix},\quad
\mathbf g_p = \begin{pmatrix}
g_{n,1}^{(p)} \\
g_{n,2}^{(p)}
\end{pmatrix}.
\label{eq:matrix_definitions_main}
\end{equation}
Here, the phenomenological parameter $J^{(p)}_{\mathrm{int}}$ represents the effective inter-film coupling. The family-resolved double layer self-energy is $\Sigma_p(\omega) = \mathbf g_p^{\,T} \left[\omega I - \Omega_p + iK_p \right]^{-1} \mathbf g_p$, yielding the total reduced double layer transmission:
\begin{equation}
S_{n,\mathrm{double}}(\omega)
=
\frac{\kappa_{c,n}}
{i(\omega-\omega_{c,n})-\kappa_{c,n}
-
i\displaystyle\sum_{p\in\mathcal O}\Sigma_p(\omega)}.
\label{eq:S_double layer_exchange_reduced_main}
\end{equation}

In the symmetric limit ($d_1=d_2$, $J_1=J_2$, $\omega_{\mathrm{SWR},1}^{(p)}=\omega_{\mathrm{SWR},2}^{(p)}$, $g_{n,1}^{(p)}=g_{n,2}^{(p)}$), the natural magnetic basis shifts to the bright and dark double layer combinations $m_{B,D}^{(p)}=(m_{1}^{(p)}\pm m_{2}^{(p)})/\sqrt{2}$. The cavity couples to the bright combination with an enhanced element $G_n^{(p)}=\sqrt{2}\,g_n^{(p)}$, while the dark combination remains decoupled. If the films are detuned ($H_1\neq H_2$ or $g_{n,1}^{(p)}\neq g_{n,2}^{(p)}$), this dark-channel cancellation is relaxed separately for each family $p$. This reduced $J\neq 0$ theory captures the essential standing-spin-wave magnon-polariton features, including mode-dependent couplings and family-resolved inter-film hybridization, without explicitly solving the full seven-region exchange-scattering problem.

\section{Results and Discussion}
\label{sec:results}

In this section, we present the physical consequences of the double layer planar-cavity theory developed in Sec.~\ref{sec:model}. We first validate the scattering formalism against the established single-film benchmark in the zero-gap limit. Next, we demonstrate geometry-controlled bright-channel enhancement in the symmetric double layer, followed by the activation of a dark-derived branch through controlled symmetry breaking. Finally, we apply the reduced $J\neq 0$ theory to reveal family-resolved bright and dark channels in the standing-spin-wave regime. For all calculations, we used yttrium iron garnet (YIG) as the magnetic material with $\mu_0 M_s = 0.175~\text{T}$, $\gamma/2\pi=28~\text{GHz/T}$, $\alpha = 3\times10^{-4}$, and $\eta=15$ \cite{Cao2015ExchangeMagnonPolaritons, sadhana2009synthesis, manuilov2009pulsed}. The cavity parameters were set to $L=46~\text{mm}$ and $\Delta=2L$.

\subsection{Validation against the single-film benchmark}
\label{subsec:results_validation}

\begin{figure}[t]
    \centering
    \includegraphics[width=1\linewidth]{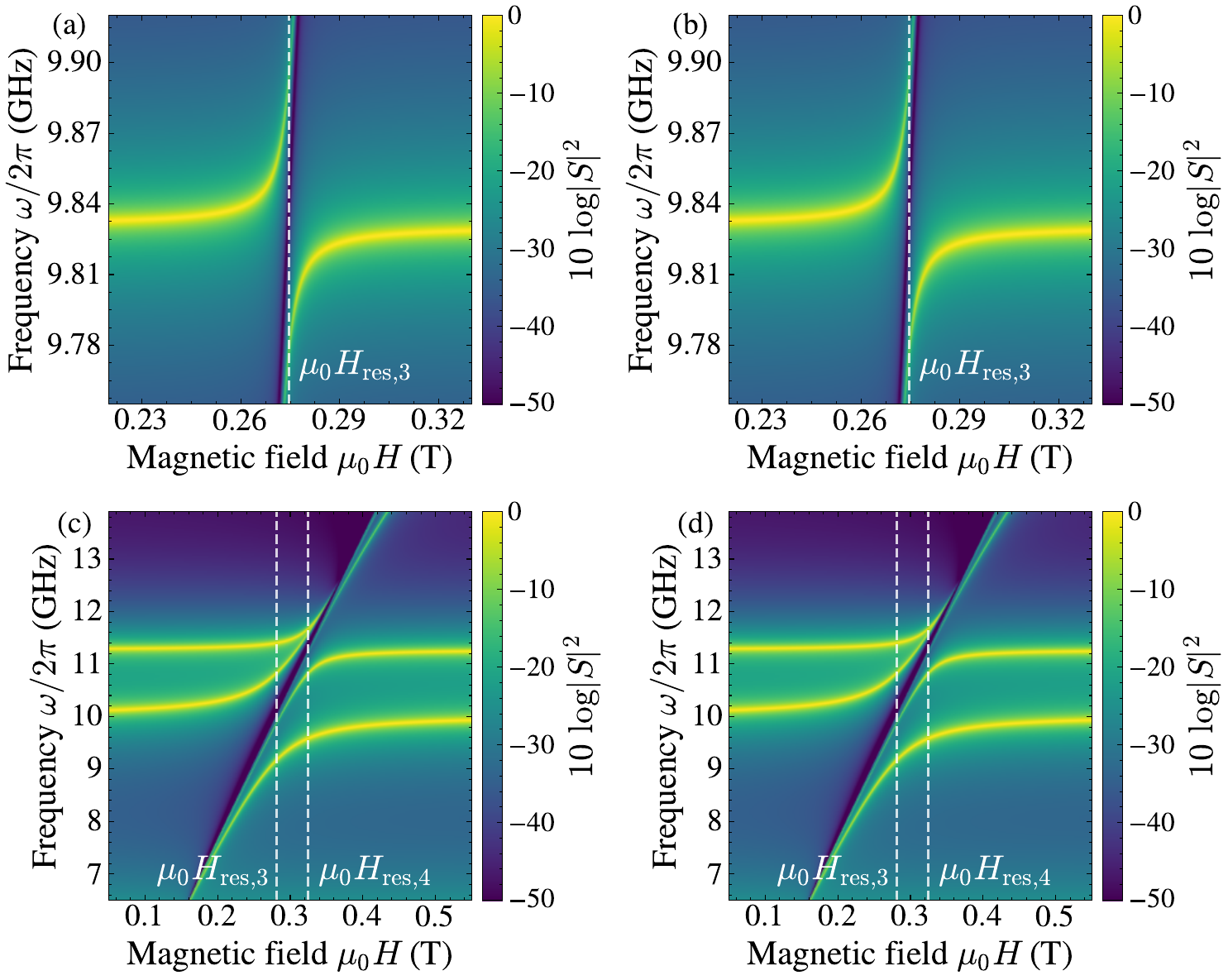}
    \caption{Validation of the double layer scattering theory against the single-film planar-cavity benchmark for a thinner film $d=5~\mu\text{m}$ (top row) and a thicker film $d=1~\text{mm}$ (bottom row).
    Panels (a) and (c) are normalized transmission maps calculated from the single-film scattering formula, Eq.~\eqref{eq:Ssingle_eq9_main}.
    Panels (b) and (d) are the corresponding maps calculated from the double layer theory for $d_1=d_2=d/2$ and $s=0$.
    The excellent agreement confirms that our double layer theory reproduces the established one-film result in the zero-gap half-thickness limit.}
    \label{fig:validation}
\end{figure}

We first verify that the double layer scattering theory reproduces the single-film planar-cavity benchmark. As formally required, the double layer transmission must collapse exactly to the one-film problem in the zero-gap half-thickness limit ($d_1=d_2=d/2$, $s=0$). Figure~\ref{fig:validation} compares the normalized transmission map obtained from the original one-film formula, Eq.~\eqref{eq:Ssingle_eq9_main}, with the double layer theory under this limiting condition, Eq.~\eqref{eq:zero_gap_identity_main}. Panels (a) and (b) show the resonance for a thin film ($d = 5\ \mu\text{m}$), which is dominated by the $n=3$ cavity photon mode at approximately $9.84$ GHz.

To confirm that the framework remains controlled even when the magnetic slab produces appreciable cavity renormalization, we also perform this comparison in a thicker-film regime. As shown in panels (c) and (d) for $d = 1$ mm, the spectral structure broadens and reveals two distinct resonance features identified as the $n=3$ mode (shifted to $\approx 10.03$ GHz) and an additional $n=4$ cavity photon mode at $\approx 11.27$ GHz.  The respective sets of maps for both the thinner and thicker films are indistinguishable. This excellent agreement confirms that the double layer construction is mathematically and numerically consistent with the established single-film theory, justifying its use as a baseline for quantifying double layer phenomena in subsequent sections.

\subsection{Symmetric double layer: geometry-controlled bright-channel enhancement}
\label{subsec:results_sym_enhancement}

In the symmetric two-film configuration ($d_1=d_2=d$, $H_1=H_2=H$, etc.), one might intuitively expect that doubling the magnetic material simply produces a larger effective coupling. However, the full scattering theory reveals a more subtle geometry-dependent behavior, showing that the enhancement of the collective bright channel depends strictly on where the two films sample the standing-wave structure of the cavity mode.

\begin{figure}[t]
    \includegraphics[width=1\linewidth]{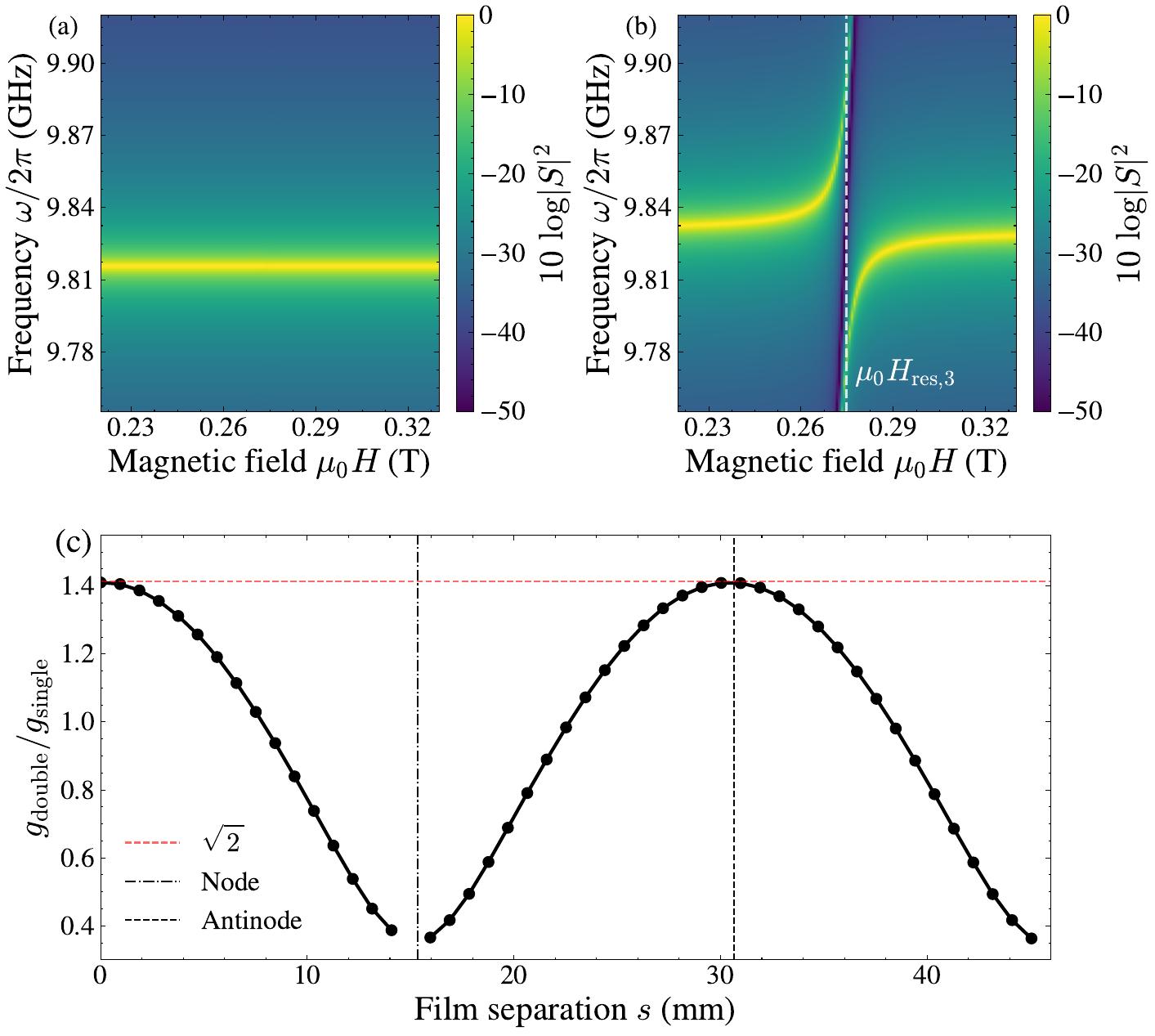}
    \caption{Normalized transmission spectra of the symmetric double layer for representative inter-film separations: (a)~$s\approx (L-2d)/3$, where the film centers approach the node-pair region (coupling suppressed), and (b)~$s\approx 2(L-2d)/3$, where the films sample the side antinodes (strong coupling recovered). Panel (c) depicts the effective coupling ratio $g_{\mathrm{double}}/g_{\mathrm{single}}$ as a function of $s$. Two antinode-compatible regimes are visible (near $s=0$ and $s\approx 2(L-2d)/3$), approaching the ideal $\sqrt{2}$ enhancement (dashed line) for the coherent bright channel. In the node-compatible region ($s\approx (L-2d)/3$), coupling is strongly suppressed. Data points are omitted in the suppressed region to highlight the enhancement peaks.}
    \label{fig:symmetric_double layer_maps}
\end{figure}

To make this explicit, we examine the normalized transmission spectra for representative inter-film separations $s$. Figures~\ref{fig:symmetric_double layer_maps}(a) and (b) show two characteristic cases. For $s \approx (L-2d)/3$, the two film centers move toward the node-pair region of the $n=3$ cavity mode, strongly reducing the overall coupling. Conversely, for $s \approx 2(L-2d)/3$, the films are positioned near the two side antinodes, and strong coupling is recovered, similar to the zero-gap case ($s = 0$). This directly demonstrates that the double layer response is governed by spatial mode overlap rather than magnetic thickness alone.

To quantify this, we extract an effective coupling from the resonance field line cuts. Defining the splitting between the two dominant hybridized peaks as $\Delta f_{\mathrm{split}}=f_{+}-f_{-}$, the effective coupling is $g_{\mathrm{eff}}=\Delta f_{\mathrm{split}}/2$. We then introduce the ratio $g_{\mathrm{double}}/g_{\mathrm{single}}$, comparing the symmetric double layer coupling to the one-film benchmark of thickness $d$. In the ideal coherent bright-channel limit, a $\sqrt{2}$-type enhancement is expected \cite{Zhang2015DarkModes,ZareRameshti2022CavityMagnonics}. 

As shown in Figure~\ref{fig:symmetric_double layer_maps}(c), this ideal enhancement is achieved in two distinct windows located near $s=0$ (central antinode) and near $s\approx 2(L-2d)/3$ (side antinodes). Outside these regions, particularly near the node pair at $s\approx (L-2d)/3$, the coupling drops significantly. The spatial origin of this modulation is clear from the film-center coordinates $x_{1,2}=(L\mp d\mp s)/2$. Beyond simply separating the films, tuning $s$ actively drives them across the standing-wave envelope. This geometric control establishes the symmetric double layer as an independent cavity-engineering platform, utilizing the internal separation $s$ as a new degree of freedom to maximize or suppress collective mode coupling \cite{Zhang2014StronglyCoupled,ZareRameshti2022CavityMagnonics,MartinezLosa2023ShapedFerromagnets}.

\subsection{Asymmetric double layer: activation of a dark-derived branch}
\label{subsec:results_asym_dark}

\begin{figure}[t]
    \centering
    \includegraphics[width=\linewidth]{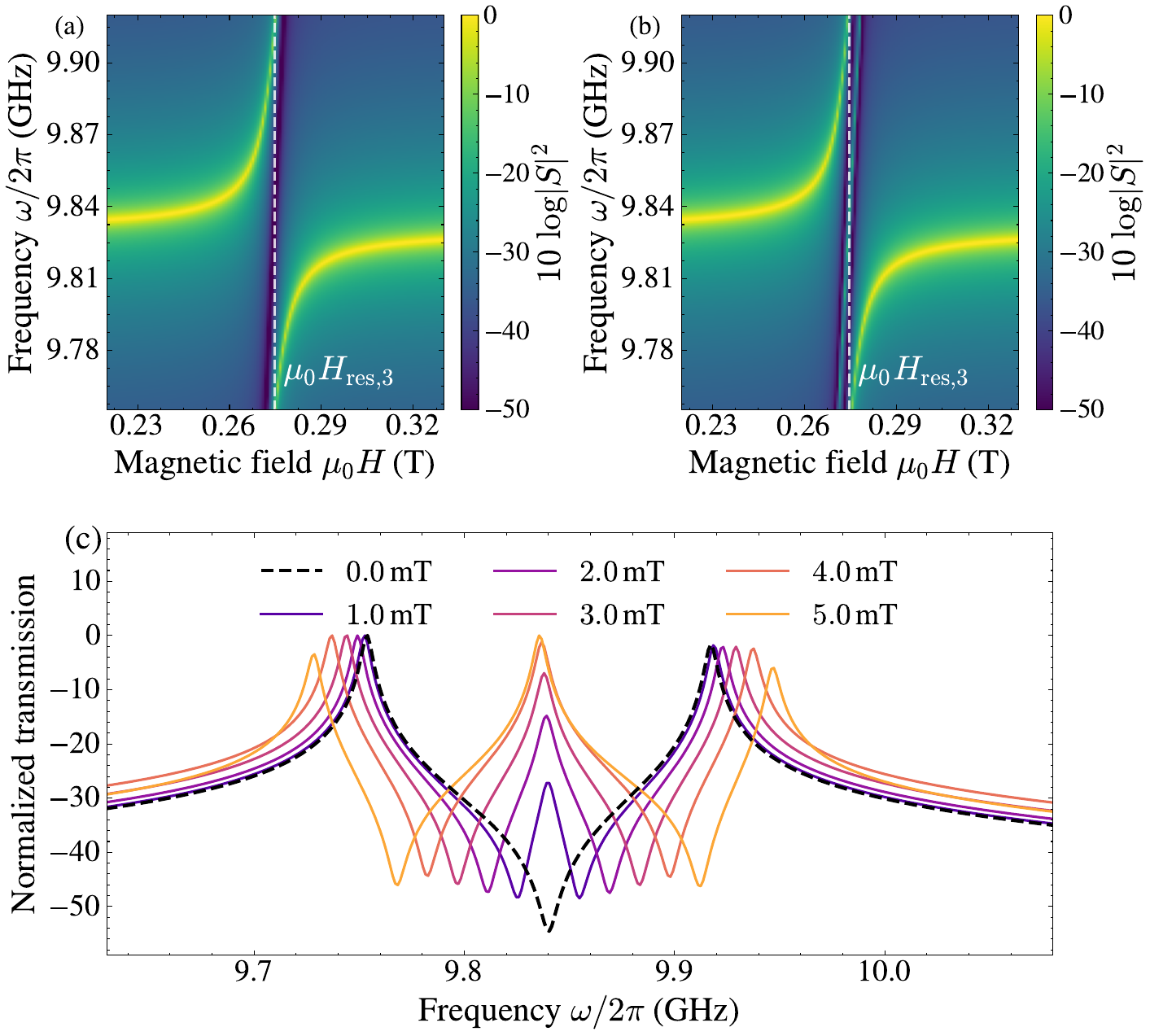}
    \caption{Full-scattering $J=0$ spectra of the double layer cavity in the symmetric and asymmetric cases.
    (a) Symmetric transmission map showing the dominant bright avoided crossing.
    (b) Asymmetric map for a field imbalance $\mu_0 \delta H=3.0~ \text{mT}$, revealing an additional weak branch. The vertical dashed lines indicate the resonance field $\mu_0 H_{\text{res,3}}$.
    (c) Representative resonance-field line cuts at the dashed lines for increasing asymmetry $\mu_0 \delta H$. As asymmetry increases, an intermediate third peak develops, signifying the cavity-activated dark-derived branch.}
    \label{fig:asym_double layer_maps}
\end{figure}

In the perfectly symmetric limit (Sec.~\ref{subsec:results_sym_enhancement}), the two films form a collective bright channel that couples strongly to the cavity, while the complementary dark channel remains decoupled \cite{Zhang2015DarkModes,ZareRameshti2022CavityMagnonics,Zhan2021BrightDarkbilayers}. However, once the two films are made inequivalent, the symmetry that protects this ideal dark state is relaxed, granting the dark-derived branch a weak but finite cavity visibility.

In the full double layer scattering theory, asymmetry enters naturally because the two films no longer share identical internal parameters. We focus on a practical control parameter, i.e., a small differential bias field:
\begin{equation}
H_1 = H + \frac{\delta H}{2},
\qquad
H_2 = H - \frac{\delta H}{2}.
\label{eq:deltaH_asym_results}
\end{equation} 
This perturbation changes the internal macrospin wave numbers and slab scattering amplitudes, breaking the left-right symmetry:
\begin{equation}
r_{\mathrm{eff}}^{L}\neq r_{\mathrm{eff}}^{R}.
\label{eq:reff_asym_results}
\end{equation}
Consequently, the cavity no longer supports a perfectly dark orthogonal channel. Instead, the dark-derived mode weakly hybridizes with the bright sector, becoming faintly visible.

Figure~\ref{fig:asym_double layer_maps} illustrates this transition. While the symmetric case [Fig.~\ref{fig:asym_double layer_maps}(a)] only displays the two main bright hybridized branches, introducing a finite asymmetry $\mu_0 \delta H$ [Fig.~\ref{fig:asym_double layer_maps}(b)] activates an additional weak branch between them.  To examine this effect directly, we analyze frequency-dependent transmission line cuts taken at the resonance field $\mu_0 H_{\text{res,3}}$ [Eq.~\eqref{eq:H_res_main}], shown in Fig.~\ref{fig:asym_double layer_maps}(c). At $\mu_0 \delta H = 0$, the spectrum is dominated by the lower and upper polariton branches of the bright FMR mode. As $\mu_0 \delta H$ increases, a third intermediate peak gradually develops between them, representing the dark-derived branch. This peak gains spectral weight with increasing asymmetry, confirming its finite cavity component. Crucially, the outer two peaks maintain a substantial bright splitting over a wide asymmetry range, allowing the activation of the dark-derived branch without immediately destroying the primary bright channel.

\begin{figure}[t]
    \centering
    \includegraphics[width=1\linewidth]{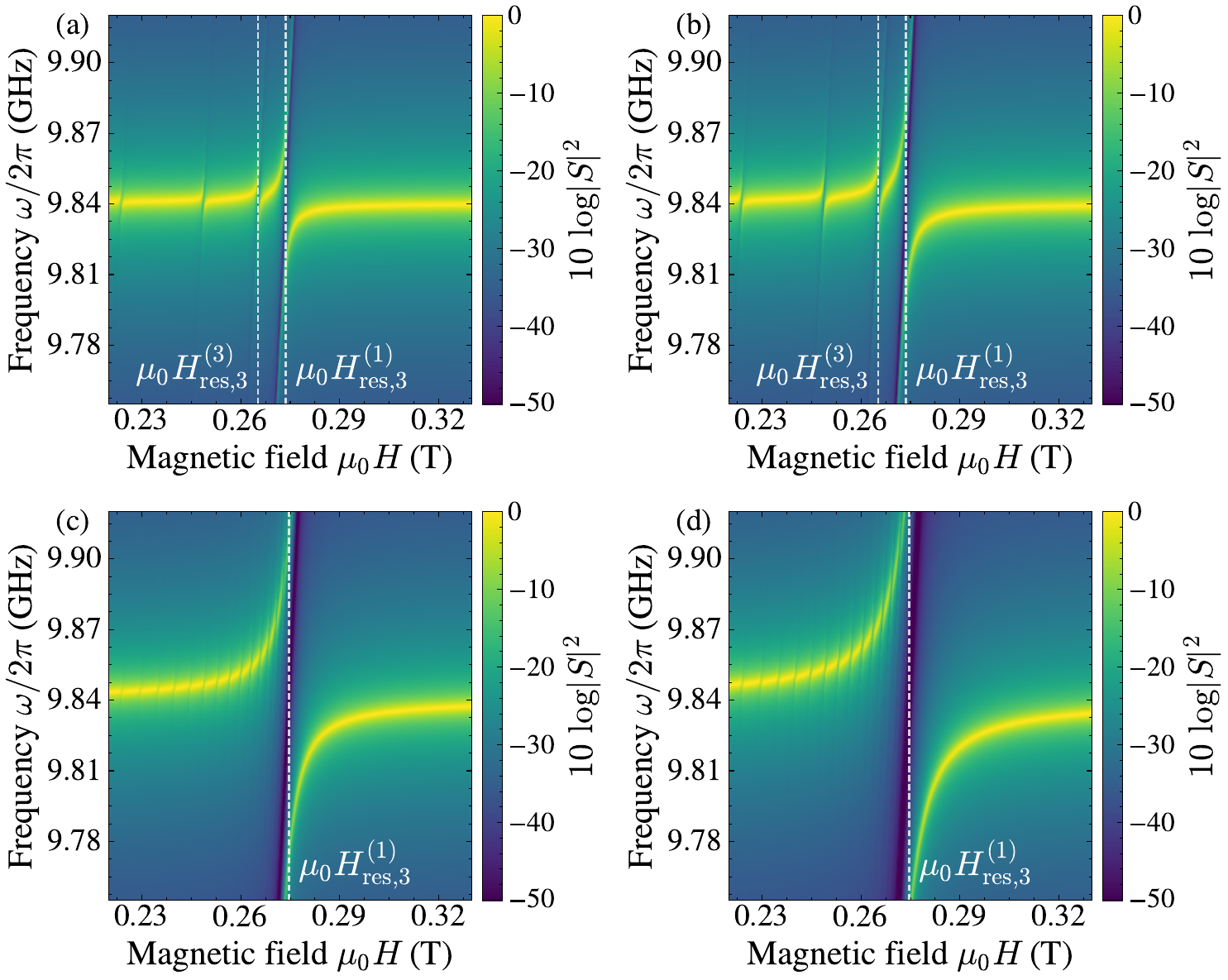}
    \caption{Comparison of the standing-spin-wave spectra between the single-film and symmetric double layer planar cavities in the exchange regime with $J =3\times10^{-16} ~\text{m}^2$ for a thinner film $d=1~\mu\text{m}$ (top row) and a thicker film $d=5~\mu\text{m}$ (bottom row) under the zero field asymmetry condition ($\mu_0 \delta H=0$).
    Panels (a) and (c) display the normalized transmission maps for a single magnetic film of thickness $d$. 
    Panels (b) and (d) show the corresponding maps for the symmetric double layer ($d_1=d_2=d$). 
    The comparison illustrates the enhanced bright-family anticrossings in the symmetric two-film geometry.}
    \label{fig:exchange_spectra}
\end{figure}

Physically, asymmetry introduces mixing between the predominantly coupled bright combination and the otherwise decoupled dark combination through unequal magnetic detunings. This activation is most effective in geometries where the bright branch is inherently strong, thus complementing the symmetric placement findings. The emergence of this weak additional branch is a genuine consequence of the same cavity-scattering theory that correctly reproduces the single-film benchmark and the symmetric double layer enhancement.

\subsection{Reduced \texorpdfstring{$J \neq 0$}{J ≠ 0} spectra}
\label{subsec:results_exchange}

We now apply the reduced multimode theory (Sec.~\ref{subsec:exchange_reduced}) to the exchange-driven regime. Here, each odd standing-spin-wave family ($p \in \{1, 3, 5, \dots\}$) forms its own double layer bright and dark channels. For identical films ($d_1=d_2=d$), we truncate the mode series at a finite $p_{\max}$ and focus on the most spectroscopically accessible families, $p=1$ and $p=3$. 

\begin{figure}[t]
    \centering
    \includegraphics[width=1\linewidth]{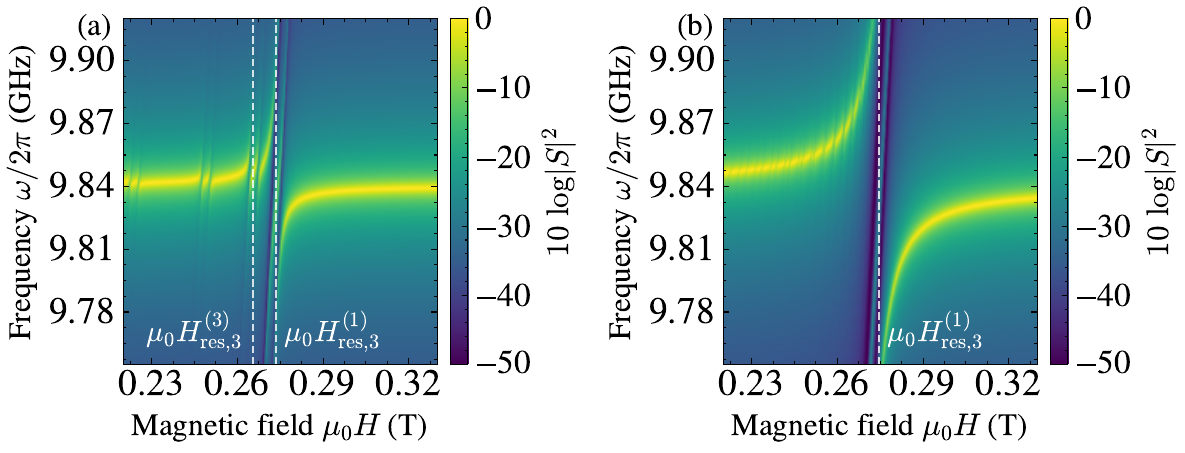}
    \caption{Normalized transmission spectra of the asymmetric double layer planar cavity for $J = 3 \times 10^{-16}~\text{m}^2$ and $\mu_0 \delta H = 3.0$~mT (cavity mode $n=3$). Panel (a) shows a thinner film $d = 1~\mu\text{m}$; panel (b) a thicker film $d = 5~\mu\text{m}$. The vertical dashed lines indicate the resonance fields $\mu_0 H_{\text{res}}^{(p)}$ [Eq.~\eqref{eq:H_res_swr}] where line cuts for Fig.~\ref{fig:exchange_family_linecuts} are taken. Symmetry breaking clearly activates the dark-derived standing-spin-wave branches.}
    \label{fig:exchange_spectra_asym}
\end{figure}

\begin{figure}[t]
    \centering
    \includegraphics[width=1\linewidth]{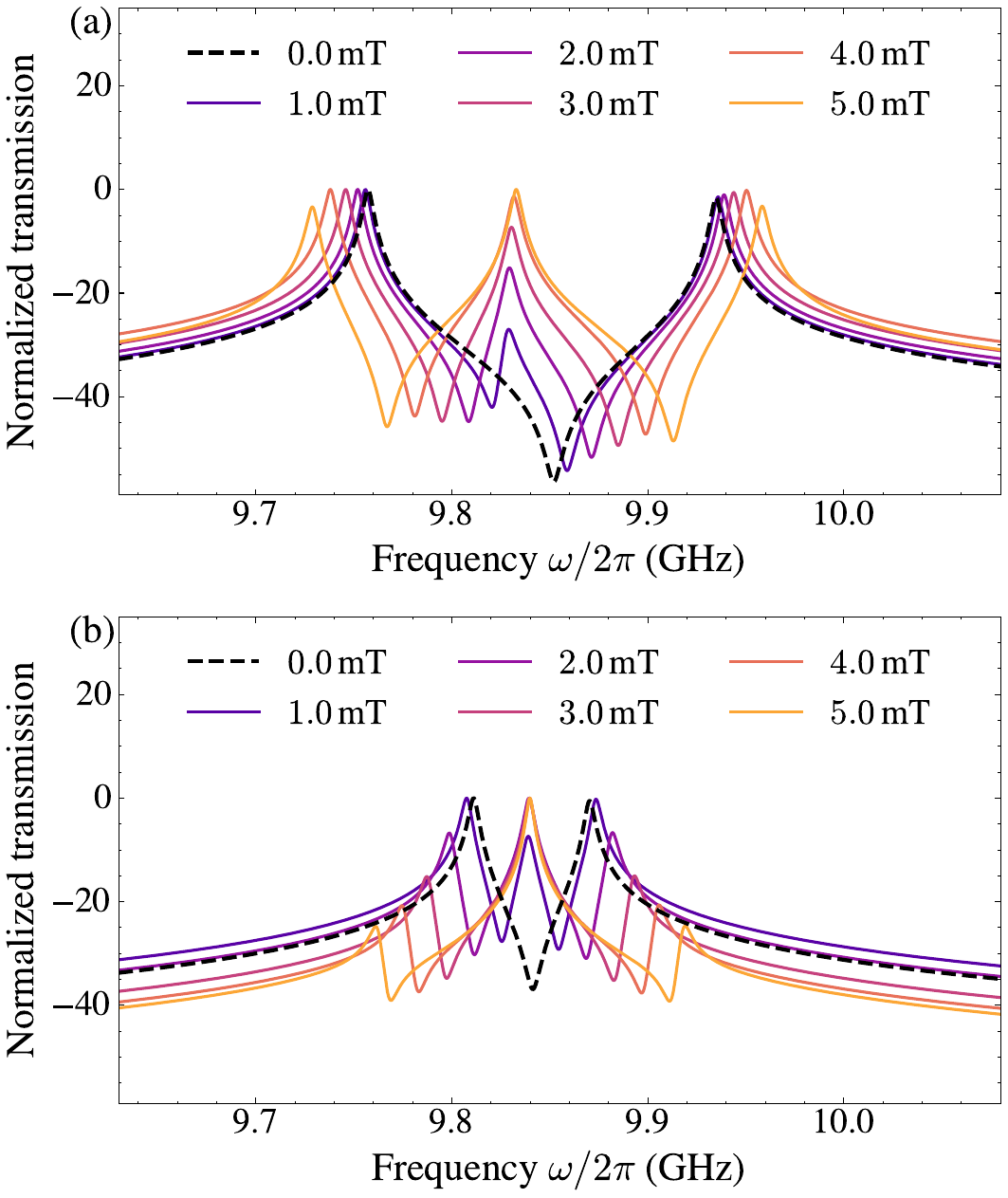}
    \caption{Representative resonance-field line cuts of the asymmetric double layer planar cavity ($J =3\times10^{-16} ~\text{m}^2$, $d_1 = d_2 = 5~\mu\text{m}$). 
    Panel (a) displays the transmission for the first odd standing-spin-wave family ($p = 1$) under increasing field asymmetry $\mu_0 \delta H$. 
    Panel (b) shows the corresponding line cuts for the third odd family ($p = 3$). 
    The progressive emergence of a middle peak illustrates the family-resolved activation of dark-derived channels. The $p=3$ dark peak is more sensitive, becoming prominent at lower $\mu_0 \delta H$ levels than $p=1$.}
    \label{fig:exchange_family_linecuts}
\end{figure}

Figures~\ref{fig:exchange_spectra}, \ref{fig:exchange_spectra_asym}, and \ref{fig:exchange_family_linecuts} summarize the reduced $J\neq 0$ spectra. We used an exchange constant $J = 3 \times 10^{-16}$~m$^2$ \cite{Serga2010YIGMagnonics} and phenomenologically introduced a family-dependent inter-layer coupling $J_{\text{int}}^{(p)} = J_{\text{int}}^{0}/p$, where $J_{\text{int}}^{0} = 12$ MHz represents the spacer $s$. The cavity targets the $n=3$ mode with $\omega_{c,3}/2\pi = 9.84$~GHz and $\kappa_{c,3}/2\pi = 1.44$~MHz \cite{Cao2015ExchangeMagnonPolaritons}. The film-cavity coupling is scaled as $g_{3,\nu}^{(p)} = g_{\text{ref}} \sqrt{d_\nu} / p$, with $g_{\text{ref}}/2\pi = 28$~MHz. Additionally, we apply a $1/p$ scaling to the interlayer coupling ($J_{\text{int}}^{(p)}$) and the film-cavity coupling strength ($g_{3,\nu}^{(p)}$) to ensure that the splitting of higher-order modes remains spectrally resolvable and does not prematurely collapse into the broad FMR region.  Compared to the single-film baseline [Figs.~\ref{fig:exchange_spectra}(a,c)], the symmetric double layer exhibits stronger bright-family anticrossings [Figs.~\ref{fig:exchange_spectra}(b,d)]. Under an asymmetric field [Figs.~\ref{fig:exchange_spectra_asym}], an additional middle feature emerges.  As discussed in Sec.~\ref{subsec:results_asym_dark}, this feature is the exchange-sector analog of the dark-derived branch, acquiring finite cavity visibility once the symmetry is broken via Eq.~\eqref{eq:deltaH_asym_results}.

To evaluate the family-resolved spectral response, in Fig.~\ref{fig:exchange_family_linecuts} we show representative line cuts extracted at $\mu_0 H_{\text{res}}^{(p)}$ [Eq.~\eqref{eq:H_res_swr}] for $d_1=d_2=5~\mu\text{m}$. Analogous to the FMR regime, the three peaks originate from the hybridized SWR-cavity system. The two outer peaks represent the lower and upper polariton branches formed by the bright SWR mode and the cavity photon.  As field asymmetry $\mu_0 \delta H$ increases, a middle peak representing the dark-derived channel emerges for both the $p=1$ [Fig.~\ref{fig:exchange_family_linecuts}(a)] and $p=3$ [Fig.~\ref{fig:exchange_family_linecuts}(b)] families. The outer peaks remain well-separated, demonstrating that a visibly activated dark-derived family can coexist with a strong bright-family anticrossing. Notably, the $p=3$ family exhibits a higher sensitivity, showing a prominent dark-derived peak at lower $\mu_0 \delta H$ levels than $p=1$. This occurs because the $p=3$ family possesses a smaller intrinsic hybridization splitting, allowing a weaker perturbation to relax the dark-state protection and transfer cavity visibility to the middle branch. 

\subsection{Discussion}
\label{sec:discussion}

The results presented above establish the double layer planar cavity as an intrinsically geometric strong-coupling platform. Rather than merely adding magnetic volume, the internal separation $s$ acts as a new degree of freedom to control spatial mode overlap, dictating whether the collective bright channel is enhanced or suppressed. Furthermore, the activation of the dark-derived branch under asymmetry emerges naturally from the cavity-wall formulation itself, proving that this phenomenon does not strictly require a reduced effective Hamiltonian to be observed.  

Several limitations of the present work should be noted. First, the exact full-scattering analysis has been carried out only in the macrospin limit ($J=0$). For the case $J\neq 0$, the spectra are obtained from a reduced multimode model. Although this model follows the established exchange-sector framework \cite{Cao2015ExchangeMagnonPolaritons}, it is not yet the full exact exchange-scattering solution. Second, the effective inter-film coupling $J_{\text{int}}^{(p)}$ is introduced phenomenologically in the reduced theory. A more microscopic treatment of how this quantity depends on film separation and standing-spin-wave family would improve the predictive capability of the model. Third, the present treatment is restricted to a one-dimensional planar cavity and therefore does not include additional features that may appear in realistic devices, such as transverse cavity modes and nonuniform demagnetizing fields. 

The limitations may suggest several directions for future work. The most immediate theoretical task is to construct the exact $J\neq 0$ bilayer scattering solver by matching the multibranch internal field expansions across all boundaries. A systematic study of direct and cavity-mediated inter-film coupling will also be important, particularly in regimes where dissipative, coherent, and non-Hermitian effects compete \cite{Grigoryan2019DissipativeSpinSpin,Harder2018LevelAttraction,Wang2019Nonreciprocity,Rao2021PerfectAbsorption}. Experimentally, planar cavities with two magnetic films and independently tunable bias fields should already be suitable for testing the main macrospin predictions of the present work, including geometry-controlled bright enhancement and weak dark-branch activation.
\section{Conclusion}
\label{sec:conclusion}

We have developed a double layer planar-cavity scattering theory in the macrospin limit ($J=0$), systematically extending the single-film framework of Cao \textit{et al.}~\cite{Cao2015ExchangeMagnonPolaritons}. By verifying the exact zero-gap half-thickness limit, we established this model as a controlled and mathematically consistent extension. Within this validated framework, we demonstrated that the symmetric double layer exhibits geometry-controlled bright-channel enhancement, where antinode-compatible placements achieve a $\sqrt{2}$-type enhancement relative to the single-film benchmark, while node-compatible placements strongly suppress the coupling. Furthermore, we showed that introducing a controlled field asymmetry activates a symmetry-protected dark state, generating an additional resonance branch without destroying the primary bright anticrossings. We then extended this picture into the exchange-driven regime ($J\neq 0$) via a reduced multimode theory, revealing that each odd standing-spin-wave family forms its own double layer bright and dark channels. The resulting spectra indicate that exchange-driven magnon-polaritons provide a significantly richer setting for collective-mode engineering than the macrospin case alone. Ultimately, this work establishes double layer planar cavities as a versatile architecture for geometry-controlled bright enhancement, weak dark-channel activation, and family-resolved standing-spin-wave hybridization.

\section*{Data and code availability}
The raw/processed data and codes required to reproduce these findings are available to download from \url{https://github.com/solihinn17/double layer-cavity-magnonics}.

\section*{CRediT authorship contribution statement}
\textbf{S.Solihin:} Data Curation, Formal analysis, Investigation, Software, Visualization,  Writing--original draft. \textbf{A.R.T.Nugraha:} Conceptualization, Formal analysis, Funding acquisition, Investigation, Methodology, Supervision, Validation, Visualization, Writing--original draft, Writing--review \& editing. \textbf{M.A.Majidi:} Funding acquisition, Project administration, Supervision, Validation, Writing--review \& editing. 

\section*{Declaration of Interests}
The authors declare that they have no known competing financial interests or personal relationships that could have appeared to influence the work reported in this paper.

\section*{Acknowledgments}
We thank QuasiLab and Mahameru BRIN for their minicluster and HPC facilities.  S.S. is supported by a research assistantship and Degree-by-Research scholarship from the BRIN Directorate for Talent Management.

\biboptions{sort&compress}
\bibliography{references}
\end{document}